\documentclass[acmsmall,nonacm]{acmart}

\acmJournal{PACMPL}
\acmVolume{1}
\acmNumber{CONF} 
\acmArticle{1}
\acmYear{2018}
\acmMonth{1}
\acmDOI{} 
\startPage{1}

\setcopyright{none}

\usepackage{booktabs}   
\usepackage{subcaption} 
                        
\usepackage[utf8]{inputenc}
\usepackage[main=english,czech]{babel}
\usepackage[T1]{fontenc}    
\usepackage{afterpage}      
\usepackage{csquotes}
\usepackage{float}
\floatstyle{ruled}
\newfloat{program}{t}{lop}
\floatname{program}{Program}
\usepackage{fancyvrb}
\usepackage{parcolumns}
\usepackage{wrapfig}
\usepackage{lipsum}

\usepackage{color}

\newcommand{\lifter}{\texttt{LiFtEr}}
\newcommand{\etal}{\textit{et al.}}

\begin{document}

\title[]{Domain-Specific Language to Encode Induction Heuristics}         


\author{Yutaka Nagashima}
\orcid{0000-0001-6693-5325}             
\affiliation{
  \position{PhD candidate}
  \department{CIIRC}             
  \institution{Czech Technical University in Prague}           
  \country{Czech Republic}                   
}
\email{Yutaka.Nagashima@cvut.cz}         
\affiliation{
  \position{PhD candidate}
  \department{Department of Computer Science}             
  \institution{University of Innsbruck}           
  \country{Austria}                   
}

\begin{abstract}
Proof assistants, such as Isabelle/HOL, 
offer tools
to facilitate inductive theorem proving. 
Isabelle experts know how to use these tools effectively; 
however, they did not have a systematic way to encode their expertise. 
To address this problem,
we present our domain-specific language, \lifter{}. 
\lifter{} allows experienced Isabelle users to encode their induction heuristics
in a style independent of any problem domain.
\lifter{}'s interpreter mechanically checks 
if a given application of induction tool matches the heuristics
specified by experienced users, 
thus systematically transferring experienced users' expertise to new Isabelle users.
\end{abstract}




\maketitle

\section{Introduction}

A noteworthy approach for proof automation in proof assistants (PAs) is the so-called hammer tools \cite{hammering}.
For example,
sledgehammer \cite{sledgehammer} exports proof goals in Isabelle/HOL \cite{isabelle}
to various external automated theorem provers (ATPs)
to exploit the state-of-the-art backend provers;
however, the discrepancies between the polymorphic higher-order logic of Isabelle/HOL
and the monomorphic first-order logic of the backend provers
severely impair sledgehammer's performance when it comes to inductive theorem proving (ITP)
\footnote{Not to be confused with \textit{interactive} theorem proving, which is a synonym of proof assistant (PA).}.

This is unfortunate for two reasons.
First, many Isabelle users chose Isabelle/HOL
precisely because its higher-order logic is 
expressive enough to specify mathematical objects and procedures involving recursion.
Second, 
induction lies at the heart of programming language semantics.
For instance, induction is often necessary for reasoning about 
recursive data-structures, such as lists and trees,
computer programs containing recursion and iteration \cite{alan1}.

Ironically, the most well-known approach, 
called waterfall \cite{waterfall}, 
for ITP was invented 
for a first-order logic,
which cannot handle induction 
without jeopardizing the soundness by introducing induction axioms. 
Jiang \etal{} took a similar approach 
and ran multiple waterfalls \cite{jiang} 
to automate ITP in HOL light \cite{hollight}.
However, when deciding induction variables,
they naively picked the first free variable with recursive type and 
left the selection of appropriate induction variables 
as future work.
To determine induction variables automatically,
we developed a proof strategy language 
\verb|PSL| and its default proof strategy, \verb|try_hard| \cite{psl}. 
\verb|PSL| tries to identify useful arguments for the \verb|induct| method by conducting a depth-first search.
However, \verb|PSL| fails to identify useful arguments
when it cannot finish a proof.

\section{Induction in Isabelle/HOL}
Isabelle offers the \verb|induct| proof method
\footnote{Proof methods are the Isar syntactic layer of LCF-style tactics.} to handle inductive problems.
For example, consider the following reverse functions, \verb|rev| and \verb|itrev|, from literature \cite{concrete_semantics}:
\begin{verbatim}
primrec rev::"'a list =>'a list" where
  "rev  []      = []"
| "rev (x # xs) = rev xs @ [x]"
fun itrev::"'a list =>'a list =>'a list" where
  "itrev  []    ys = ys"
| "itrev (x#xs) ys = itrev xs (x#ys)"
\end{verbatim}
\noindent
where \verb|#| is the list constructor, 
and \verb|@| appends two lists into one.
One can prove the equivalence of these reverse functions
in multiple ways using the \verb|induct| method:
\begin{verbatim}
lemma prf1:"itrev xs ys = rev xs @ ys" apply(induct xs arbitrary: ys) by auto
lemma prf2:"itrev xs ys = rev xs @ ys" apply(induct xs ys rule:itrev.induct) by auto
\end{verbatim}

\noindent
\verb|prf1| applies structural induction on \verb|xs|
while generalizing \verb|ys| before applying induction
by passing \verb|ys| to the \texttt{arbitrary} field.
On the other hand, \verb|prf2| applies functional induction on \verb|itrev| by passing an auxiliary lemma, \verb|itrev.induct|,
to the \texttt{rule} field.

There are other lesser-known techniques to handle difficult inductive problems using the \verb|induct| method, 
and sometimes users have to 
develop useful auxiliary lemmas manually;
however, for most cases
the problem of how to apply induction 
boils down to the the following three questions:

\begin{itemize}
    \item On which terms do we apply induction?
    \item Which variables do we generalize?
    \item Which rule do we use for functional induction?
\end{itemize}

Isabelle experts often apply induction heuristics
to answer such questions and decide what arguments to pass to
the \verb|induct| method;
however, they did not have a systematic way to encode such heuristics,
which made it difficult for new users to learn
how to apply induction effectively.

\section{\lifter{}: Language to Encode Induction Heuristics}
We address this problem with our domain-specific language, \lifter{}.
\lifter{} allows experienced Isabelle users to encode their induction heuristics
in a style independent of problem domains.
\lifter{}'s interpreter mechanically checks 
if a given application of induction is compatible with the induction heuristics
written by experienced users.
%

We designed \lifter{} to encode induction heuristics as assertions on invocations of the \verb|induct| method in Isabelle.
An assertion written in \lifter{} takes a triple of 
a proof goal at hand, its underlying proof state, 
and the arguments passed to the \verb|induct| method to prove the goal.
When one applies a \lifter{} assertion to an invocation of the \verb|induct| method,
\lifter{}'s interpreter returns a boolean value as the result of the assertion applied to the triple.

The goal of a \lifter{} programmer is to write
assertions that implement reliable heuristics.
A heuristic encoded as a \lifter{} assertion is reliable 
when it satisfies the following two properties:
first, 
the \lifter{} interpreter is likely to evaluate 
the assertion to \verb|true|
when the arguments of the \verb|induct| method are
appropriate for the given proof goal.
Second, the interpreter is likely to evaluate the assertion to \verb|false|
when the arguments are inappropriate for the goal.

\section{\lifter{} through an Example}
The following is an example assertion written in \lifter{}:
\begin{verbatim}
 Some_Rule (Rule 1, True)
Imply
 Some_Rule (Rule 1,
  Some_Trm (Trm 1,
   Some_Trm_Occ_Of (Trm_Occ 1, Trm 1,
     (Rule 1 Is_Rule_Of Trm_Occ 1)
    And 
     Is_Recursive_Cnst (Trm_Occ 1)
    And
     (All_Ind (Trm 2,
      (Some_Trm_Occ_Of (Trm_Occ 2, Trm 2,
        Some_Numb (Numb 1,
          Is_Nth_Arg_Of (Trm_Occ 2, Numb 1, Trm_Occ 1)
         And
          (Trm 2 Is_Nth_Ind Numb 1)))))))));
\end{verbatim}
\noindent
where functions starting with the string \verb|Some| are existential quantifiers,
while those starting with \verb|All| are universal quantifiers.
Quantifiers ending with \verb|Trm| describe sub-terms appearing
in the proof goal at hand, 
whereas those ending with \texttt{Trm\_Occ} describe sub-term \textit{occurrences} of such sub-terms.
It is important to distinguish terms and term occurrences
because the \verb|induct| method in Isabelle/HOL only allows its users to specify induction terms
but it does not allow us to specify on which occurrences of such terms we intend to apply induction.
Quantifiers ending with \texttt{Rule} describe auxiliary lemmas
passed to the \verb|rule| field in the \verb|induct| method,
while those ending with \texttt{Ind} describe induction terms 
specified in a given invocation of the \verb|induct| method.

\texttt{Is\_Rule\_Of}, \texttt{Is\_Recursive\_Cnst}, \texttt{Is\_Nth\_Ind}, and \texttt{Is\_Nth\_Arg\_Of} 
are atomic assertions.
For example,
\texttt{Rule n Is\_Rule\_Of Trm\_Occ m} returns true if
\texttt{Rule n} was derived by Isabelle automatically
at the time of defining the term that has an occurrence, \texttt{Trm\_Occ m}.
And \texttt{Trm n Is\_Nth\_Ind Numb m} checks if \texttt{Trm n} is 
the \texttt{(Numb m)}th induction term 
in the given invocation of the \verb|induct| method.
Importantly, some atomic assertions check 
not only the syntactic structure of the given proof goal 
and the arguments of the \verb|induct| method,
but they also investigate the semantics of terms or 
types appearing in the goal.
For example,
\texttt{Is\_Recursive\_Cnst} checks
if a given constant is defined recursively or not,
by looking up the definition of the constant stored in the proof state.
As a whole this example assertion checks if the following holds:

\begin{displayquote}
if there exists a rule, \texttt{Rule 1}, 
in the \texttt{rule} field of the \verb|induct| method,
then there exists a term \texttt{Trm 1} with an occurrence \texttt{Trm\_Occ 1}, 
such that
\texttt{Rule 1} is derived by Isabelle when defining \texttt{Trm 1},
where \texttt{Trm 1} is defined recursively, and
for all induction terms \texttt{Trm 2},
there exists an occurrence \texttt{Trm\_Occ 2} of \texttt{Trm 2} such that,
there exists a number \texttt{Numb 1}, such that
\verb|Trm_Occ 2| is the \texttt{(Numb 1)}th argument of \verb|Trm_Occ 1| and that
\verb|Trm 2| is the \texttt{(Numb 1)}th induction term passed to the \verb|induct| method
\footnote{
Note that \texttt{1} in \texttt{Numb 1} is merely the identifier
of this variable, 
and the value of \texttt{Numb 1} can be a number that is not $1$.}.
\end{displayquote}

This assertion is useful when judging invocations of 
the \verb|induct| method with an argument to the \verb|rule| field.
For example, \lifter{} can confirm that \verb|prf2| is promising
based on this assertion 
because there is an argument, \texttt{itrev.induct}, in the \verb|rule| field, where
\verb|itrev| is defined recursively,
and the occurrence of its related term, \verb|itrev|, 
in the proof goal takes
all the induction terms, \verb|xs| and \verb|ys|, 
as its arguments in the same order.

\section{Discussion and Future Work}
We presented \lifter{} and its example assertion.
\lifter{} is a domain-specific language in the sense that
we developed \lifter{} to encode induction heuristics;
however, heuristics written in \lifter{} are usually not specific to any problem domain,
because \lifter{}'s language construct is not specific to
any variables names, types, or constants.
This absence encourages \lifter{} users to 
encode heuristics that are not specific to any problem domains but
are applicable to many domains.

To the best of our knowledge, \lifter{} is the first domain-specific language
that allows us to encode induction heuristics as programs.
We released a working prototype of the \lifter{} interpreter and 
six example assertions at GitHub \cite{GitHub}. 
And a more comprehensive explanation of \lifter{}'s grammar is provided in our paper draft \cite{lifter}.
We continue writing assertions in \lifter{} and
use them as a feature extractor that distills the essence of 
promising applications of induction
from existing large proof corpora, 
such as the Archive of Formal Proofs \cite{AFP},
so that we can fully automate ITP using supervised machine learning in future. 

\newpage
\begin{acks}                            
This work was supported by the European Regional Development Fund under the project 
AI \& Reasoning (reg. no.CZ.02.1.01/0.0/0.0/15\_003/0000466).
\end{acks}

\bibliography{bibfile}


\begin{thebibliography}{12}


\ifx \showCODEN    \undefined \def \showCODEN     #1{\unskip}     \fi
\ifx \showDOI      \undefined \def \showDOI       #1{#1}\fi
\ifx \showISBNx    \undefined \def \showISBNx     #1{\unskip}     \fi
\ifx \showISBNxiii \undefined \def \showISBNxiii  #1{\unskip}     \fi
\ifx \showISSN     \undefined \def \showISSN      #1{\unskip}     \fi
\ifx \showLCCN     \undefined \def \showLCCN      #1{\unskip}     \fi
\ifx \shownote     \undefined \def \shownote      #1{#1}          \fi
\ifx \showarticletitle \undefined \def \showarticletitle #1{#1}   \fi
\ifx \showURL      \undefined \def \showURL       {\relax}        \fi
\providecommand\bibfield[2]{#2}
\providecommand\bibinfo[2]{#2}
\providecommand\natexlab[1]{#1}
\providecommand\showeprint[2][]{arXiv:#2}

\bibitem[\protect\citeauthoryear{Blanchette, Kaliszyk, Paulson, and
  Urban}{Blanchette et~al\mbox{.}}{2016}]%
        {hammering}
\bibfield{author}{\bibinfo{person}{Jasmin Blanchette}, \bibinfo{person}{Cezary
  Kaliszyk}, \bibinfo{person}{Lawrence Paulson}, {and} \bibinfo{person}{Josef
  Urban}.} \bibinfo{year}{2016}\natexlab{}.
\newblock \showarticletitle{Hammering towards QED}.
\newblock \bibinfo{journal}{\emph{Journal of Formalized Reasoning}}
  \bibinfo{volume}{9}, \bibinfo{number}{1} (\bibinfo{year}{2016}),
  \bibinfo{pages}{101--148}.
\newblock
\showISSN{1972-5787}
\urldef\tempurl%
\url{https://doi.org/10.6092/issn.1972-5787/4593}
\showDOI{\tempurl}


\bibitem[\protect\citeauthoryear{Blanchette, B{\"{o}}hme, and
  Paulson}{Blanchette et~al\mbox{.}}{2011}]%
        {sledgehammer}
\bibfield{author}{\bibinfo{person}{Jasmin~Christian Blanchette},
  \bibinfo{person}{Sascha B{\"{o}}hme}, {and} \bibinfo{person}{Lawrence~C.
  Paulson}.} \bibinfo{year}{2011}\natexlab{}.
\newblock \showarticletitle{Extending Sledgehammer with {SMT} Solvers}. In
  \bibinfo{booktitle}{\emph{Automated Deduction - {CADE-23} - 23rd
  International Conference on Automated Deduction, Wroclaw, Poland, July 31 -
  August 5, 2011. Proceedings}} \emph{(\bibinfo{series}{Lecture Notes in
  Computer Science})}, \bibfield{editor}{\bibinfo{person}{Nikolaj Bj{\o}rner}
  {and} \bibinfo{person}{Viorica Sofronie{-}Stokkermans}} (Eds.),
  Vol.~\bibinfo{volume}{6803}. \bibinfo{publisher}{Springer},
  \bibinfo{pages}{116--130}.
\newblock
\showISBNx{978-3-642-22437-9}
\urldef\tempurl%
\url{https://doi.org/10.1007/978-3-642-22438-6}
\showDOI{\tempurl}


\bibitem[\protect\citeauthoryear{Bundy}{Bundy}{2001}]%
        {alan1}
\bibfield{author}{\bibinfo{person}{Alan Bundy}.}
  \bibinfo{year}{2001}\natexlab{}.
\newblock \showarticletitle{The Automation of Proof by Mathematical Induction}.
\newblock In \bibinfo{booktitle}{\emph{Handbook of Automated Reasoning (in 2
  volumes)}}, \bibfield{editor}{\bibinfo{person}{John~Alan Robinson} {and}
  \bibinfo{person}{Andrei Voronkov}} (Eds.). \bibinfo{publisher}{Elsevier and
  {MIT} Press}, \bibinfo{pages}{845--911}.
\newblock
\showISBNx{0-444-50813-9}


\bibitem[\protect\citeauthoryear{Harrison}{Harrison}{1996}]%
        {hollight}
\bibfield{author}{\bibinfo{person}{John Harrison}.}
  \bibinfo{year}{1996}\natexlab{}.
\newblock \showarticletitle{{HOL} Light: {A} Tutorial Introduction}. In
  \bibinfo{booktitle}{\emph{Formal Methods in Computer-Aided Design, First
  International Conference, {FMCAD} '96, Palo Alto, California, USA, November
  6-8, 1996, Proceedings}} \emph{(\bibinfo{series}{Lecture Notes in Computer
  Science})}, \bibfield{editor}{\bibinfo{person}{Mandayam~K. Srivas} {and}
  \bibinfo{person}{Albert~John Camilleri}} (Eds.), Vol.~\bibinfo{volume}{1166}.
  \bibinfo{publisher}{Springer}, \bibinfo{pages}{265--269}.
\newblock
\showISBNx{3-540-61937-2}
\urldef\tempurl%
\url{https://doi.org/10.1007/BFb0031814}
\showDOI{\tempurl}


\bibitem[\protect\citeauthoryear{Jiang, Papapanagiotou, and Fleuriot}{Jiang
  et~al\mbox{.}}{2018}]%
        {jiang}
\bibfield{author}{\bibinfo{person}{Yaqing Jiang}, \bibinfo{person}{Petros
  Papapanagiotou}, {and} \bibinfo{person}{Jacques~D. Fleuriot}.}
  \bibinfo{year}{2018}\natexlab{}.
\newblock \showarticletitle{Machine Learning for Inductive Theorem Proving}. In
  \bibinfo{booktitle}{\emph{Artificial Intelligence and Symbolic Computation -
  13th International Conference, {AISC} 2018, Suzhou, China, September 16-19,
  2018, Proceedings}} \emph{(\bibinfo{series}{Lecture Notes in Computer
  Science})}, \bibfield{editor}{\bibinfo{person}{Jacques~D. Fleuriot},
  \bibinfo{person}{Dongming Wang}, {and} \bibinfo{person}{Jacques Calmet}}
  (Eds.), Vol.~\bibinfo{volume}{11110}. \bibinfo{publisher}{Springer},
  \bibinfo{pages}{87--103}.
\newblock
\showISBNx{978-3-319-99956-2}
\urldef\tempurl%
\url{https://doi.org/10.1007/978-3-319-99957-9\_6}
\showDOI{\tempurl}


\bibitem[\protect\citeauthoryear{Klein, Nipkow, Paulson, and Thiemann}{Klein
  et~al\mbox{.}}{2004}]%
        {AFP}
\bibfield{author}{\bibinfo{person}{Gerwin Klein}, \bibinfo{person}{Tobias
  Nipkow}, \bibinfo{person}{Larry Paulson}, {and} \bibinfo{person}{Rene
  Thiemann}.} \bibinfo{year}{2004}\natexlab{}.
\newblock \bibinfo{booktitle}{}.
\newblock
\showISSN{2150-914x}
\urldef\tempurl%
\url{https://www.isa-afp.org/}
\showURL{%
\tempurl}


\bibitem[\protect\citeauthoryear{Moore}{Moore}{1973}]%
        {waterfall}
\bibfield{author}{\bibinfo{person}{J.~Strother Moore}.}
  \bibinfo{year}{1973}\natexlab{}.
\newblock \emph{\bibinfo{title}{Computational logic : structure sharing and
  proof of program properties}}.
\newblock \bibinfo{thesistype}{Ph.D. Dissertation}. \bibinfo{school}{University
  of Edinburgh, {UK}}.
\newblock
\urldef\tempurl%
\url{http://hdl.handle.net/1842/2245}
\showURL{%
\tempurl}


\bibitem[\protect\citeauthoryear{Nagashima}{Nagashima}{2019}]%
        {lifter}
\bibfield{author}{\bibinfo{person}{Yutaka Nagashima}.}
  \bibinfo{year}{2019}\natexlab{}.
\newblock \bibinfo{title}{LiFtEr: Language to Encode Induction Heuristics for
  Isabelle/HOL}.
\newblock
\newblock
\showeprint[arxiv]{1906.08084}


\bibitem[\protect\citeauthoryear{Nagashima et~al\mbox{.}}{Nagashima
  et~al\mbox{.}}{2019}]%
        {GitHub}
\bibfield{author}{\bibinfo{person}{Yutaka Nagashima} {et~al\mbox{.}}}
  \bibinfo{year}{2019}\natexlab{}.
\newblock \bibinfo{title}{data61/PSL}.
\newblock
\newblock
\urldef\tempurl%
\url{https://github.com/data61/PSL/releases/tag/v0.1.3-alpha}
\showURL{%
\tempurl}


\bibitem[\protect\citeauthoryear{Nagashima and Kumar}{Nagashima and
  Kumar}{2017}]%
        {psl}
\bibfield{author}{\bibinfo{person}{Yutaka Nagashima} {and}
  \bibinfo{person}{Ramana Kumar}.} \bibinfo{year}{2017}\natexlab{}.
\newblock \showarticletitle{A Proof Strategy Language and Proof Script
  Generation for Isabelle/HOL}. In \bibinfo{booktitle}{\emph{Automated
  Deduction - {CADE} 26 - 26th International Conference on Automated Deduction,
  Gothenburg, Sweden, August 6-11, 2017, Proceedings}}
  \emph{(\bibinfo{series}{Lecture Notes in Computer Science})},
  \bibfield{editor}{\bibinfo{person}{Leonardo de~Moura}} (Ed.),
  Vol.~\bibinfo{volume}{10395}. \bibinfo{publisher}{Springer},
  \bibinfo{pages}{528--545}.
\newblock
\urldef\tempurl%
\url{https://doi.org/10.1007/978-3-319-63046-5\_32}
\showDOI{\tempurl}


\bibitem[\protect\citeauthoryear{Nipkow and Klein}{Nipkow and Klein}{2014}]%
        {concrete_semantics}
\bibfield{author}{\bibinfo{person}{Tobias Nipkow} {and} \bibinfo{person}{Gerwin
  Klein}.} \bibinfo{year}{2014}\natexlab{}.
\newblock \bibinfo{booktitle}{\emph{Concrete Semantics - With Isabelle/HOL}}.
\newblock \bibinfo{publisher}{Springer}.
\newblock
\showISBNx{978-3-319-10541-3}
\urldef\tempurl%
\url{https://doi.org/10.1007/978-3-319-10542-0}
\showDOI{\tempurl}


\bibitem[\protect\citeauthoryear{Nipkow, Paulson, and Wenzel}{Nipkow
  et~al\mbox{.}}{2002}]%
        {isabelle}
\bibfield{author}{\bibinfo{person}{Tobias Nipkow}, \bibinfo{person}{Lawrence~C.
  Paulson}, {and} \bibinfo{person}{Markus Wenzel}.}
  \bibinfo{year}{2002}\natexlab{}.
\newblock \bibinfo{booktitle}{\emph{Isabelle/HOL - a proof assistant for
  higher-order logic}}. \bibinfo{series}{Lecture Notes in Computer Science},
  Vol.~\bibinfo{volume}{2283}.
\newblock \bibinfo{publisher}{Springer}.
\newblock
\showISBNx{3-540-43376-7}


\end{thebibliography}



\end{document}